\documentstyle[aps,twocolumn,epsfig]{revtex}
\draft
\title{Analytic expression for Taylor--Couette stability boundary}
\author{Alexander Esser and Siegfried Grossmann}
\address{Fachbereich Physik der Philipps--Universit\"at, Renthof 6, D-35032, 
Marburg, Germany}
\date{\today}
\narrowtext 

\begin{document}

\maketitle
 
\begin{abstract}

We analyze the mechanism that determines the boundary of stability in
Taylor--Couette flow.
By simple physical argument
we derive an analytic expression to approximate the
stability line for all radius ratios and all speed ratios, for co- and
counterrotating cylinders.
The expression includes viscosity and so generalizes Rayleigh's criterion.
We achieve agreement with linear stability theory
and with experiments in the whole parameter space.
Explicit formulae are given for limiting cases.

\end{abstract}

\pacs{PACS 47.15.Fe, 47.20.Gv}
\preprint{}


\noindent\\{\bf 1. Introduction}\\\smallskip

The Taylor--Couette system has been studied for more than 100 years, 
beginning with the first experiments by Couette~\cite{CouMM:1888},
Mallock~\cite{MalA:1888}, and the pioneering work by Taylor~\cite{TayGI:23};
for a recent review see Tagg~\cite{TagR:94}.

We here focus on the question when and why the laminar flow becomes unstable.
This can and has been evaluated with linear stability theory
\cite{TayGI:23,CHA:61,WalJ_TsaS_DPrRC:64,JOS-1:76,DRAR:81,GebTh_Gro:93},
and was measured for a variety of radius ratios
\cite{TayGI:23,DonRJ_FulD:60,ColD:65,SnyHA:68b,AndCD_LiuSS_SwiHL:86}.
The aim of this note is to elucidate the physical mechanism and derive a
simple approximate analytical formula for the stability boundary in the whole
parameter space.

Earlier attempts of this kind have been made by
Donnelly and Fultz~\cite{DonRJ_FulD:60}, Coles~\cite{ColD:67}, 
Joseph~\cite{JOS-1:76}, and
Eckhardt and Yao~\cite{EckB_YaoD:95}.
However, these authors did not consider all relevant aspects.
Those include
the smooth crossover from a solid boundary (for corotation)
to a free one (for sufficiently strong counterrotation)
and the localization of the initial perturbation in the center of
the relevant gap.
We elaborate that in Sects.~2--4.

We consider an incompressible fluid with kinematic viscosity $\nu$ flowing 
between two concentric cylinders with inner radius $r_1$ and outer radius 
$r_2$, and angular velocities $\omega_1>0$ and $\omega_2$, positive for
corotation and negative for counterrotation.
This fluid flow is characterized by the Reynolds numbers 
	$R_i = \omega_i r_i d/\nu$ ($i=1,2$),
where $d=r_2-r_1$ is the gap size between the cylinders.

The well-known \cite{CHA:61,LL-6-E} laminar flow
is purely azimuthal and has the velocity profile
\begin{eqnarray}\label{eUlam}\label{eABDef}
	u_\varphi (r) & = & Ar + \frac{B}{r}
	\quad \mbox{with} \\
	A & = & - \frac{\eta^2\omega_1 - \omega_2}{1 - \eta^2} ,\quad
	B =	\frac{\omega_1 - \omega_2}{1 - \eta^2} \, r_1^2 . \nonumber
\end{eqnarray}
The radius ratio $\eta = r_1/r_2$ characterizes the geometry of the system.
For counterrotation a nodal surface exists where $u_\varphi=0$,
\begin{equation}\label{eNodalDef}
	r_n = r_2 \, \sqrt{\frac{R_1 - \eta R_2}{R_1 - \eta^{-1}R_2}} ,
	\quad R_2 \le 0 .
\end{equation}

\bigskip\noindent\\{\bf 2. Derivation of the stability criterion}\\\smallskip

Let us start with Rayleigh's stability argument for inviscid
flow~\cite{RayL:16}:
Consider two neighbouring fluid rings at $r$ and $r+\delta r$ of the
laminar flow.
Stability is guaranteed if their hypothetical exchange by virtual motion
through the surrounding flow costs kinetic energy.
Neglecting momentum loss by viscosity, the angular momenta per
mass $L(r)=r u_\varphi (r)$ of the rings are conserved,
and with their kinetic energy per mass being $u_\varphi^2 /2$
we arrive at Rayleigh's famous stability criterion $L^2(r)^\prime >0$
when considering the energy balance
\begin{eqnarray*}
	-\delta E_{\text{kin}} & = &
	\frac{1}{2} \Bigl( \frac{L(r+\delta r)}{r} \Bigr)^2 
	- \frac{1}{2} \Bigl( \frac{L(r+\delta r)}{r+\delta r} \Bigr)^2 \\
	&+ & \frac{1}{2} \Bigl( \frac{L(r)}{r+\delta r} \Bigr)^2
	- \frac{1}{2} \Bigl( \frac{L(r)}{r} \Bigr)^2 
	= \frac{L^2(r)^\prime}{r^3} \, (\delta r)^2 > 0 .
\end{eqnarray*}
$\delta r$ has been choosen infinitesimal and the prime denotes the 
$r$-derivative.
Stability is predicted for corotation if $R_1 < R_2/\eta$, and
in the case of counterrotation for $R_1=0$ only.

Now we take into account viscosity.
During the motion to exchange the two fluid rings as described above
they loose angular momentum and thus kinetic energy by dissipation.
The frictional energy loss per mass is given by the viscous force per mass
times the shift $\delta r$,
\begin{displaymath}
	\delta E_{\text{vis}}
	\sim \bigl( \frac{\nu}{\ell^2} \, \delta u_r \bigr) \delta r .
\end{displaymath}
$\ell$ denotes the typical diameter of the ring (in radial as
well as in axial direction).
	$\delta u_r = \delta r / \delta t$
is the radial velocity of the shift which takes place within time $\delta t$.
The `$\sim$' symbol means `up to a factor of order unity'.
For instability, 
$\delta E_{\text{kin}}$ has to exceed $\delta E_{\text{vis}}$.
Its lower limit is related to the
upper limit for $\delta t$, which is set by the diffusive time scale,
on which the rings loose angular momentum,
	$\delta t \sim \ell^2/\nu$.
We now obtain from $\delta E_{\text{kin}} \sim \delta E_{\text{vis}}$
\begin{equation}\label{eDissipation}
	-\frac{L^2(r)^\prime}{r^3} \, (\delta r)^2
	\sim \frac{\nu^2}{\ell^4} \, (\delta r)^2
\end{equation}
as a relation which characterizes the stability boundary,
or, after inserting the profil (\ref{eUlam}), i.e., $L=A r^2+B$,
\begin{equation}\label{eMain1}
	\Bigl[ \frac{2}{1+\eta} \, (\eta R_1 - R_2) \Bigr]^2 \,
	\Bigl[ \frac{r_n^2}{r^2} - 1 \Bigr] 
	\sim
	\Bigl( \frac{d}{\ell} \Bigr)^4 .
\end{equation}
Here $r_n^2=-B/A$, which for negative $R_2$ coincides with the
squared position of the nodal surface.

In contrast to the inviscid case
the timescale for the motion
and the size of the rings enter here;
the expression on the rhs. of (\ref{eDissipation}) has also been used 
in~\cite{ColD:67}.
As we will see soon, the radial position of the rings is important as well.

$\ell$ is some small but finite fraction of the part of the
gap which is susceptible to instability.
For positive $R_2$ this is the whole gap between the cylinders,
therefore we write $\ell = \alpha d$
with $\alpha$ a constant considerably smaller than $1$~\cite{ColD:67}.
We shall determine it by comparison with linear stability analysis.

In the counterrotating case $L^2(r)$ decreases in the range from the inner
cylinder up to the nodal surface only.
We therefore compare $\ell$ with $d_n = r_n - r_1$,
which is an old idea, cf also \cite{DonRJ_FulD:60}.
We further take into account that the flow patterns appearing at the 
instability
line extend beyond the nodal surface~\cite{TayGI:23}.
This may be understood as a change in the type of boundary from
solid (for $R_2 \ge 0$) to sort of free (for $R_2 < 0$).
The simplest way of describing this crossover is
\begin{equation}
	\label{eEllDef}	\label{eDeltaDef}
	\ell = \alpha d \, \Delta (a\frac{d_n}{d}) 
	\quad\mbox{with}\quad
	\Delta (x) = \left\{
		\begin{array}{rcl}
			x &	, & \text{if}\quad x<1\\
			1 & , & \text{if}\quad x\ge1
		\end{array}
	\right. .
\end{equation}
Thus, $\ell=\alpha a d_n$
for $R_2 \le R_2^\ast$, where $R_2^\ast$ is determined by
the condition $a d_n/d=1$.
If $a>1$ we get $R_2^\ast<0$.
This delay in the decrease of $\ell$ when $R_2$ becomes negative
can also be seen in Fig.~12 of Ref.~\cite{TayGI:23} for the
axial spacing of vortices.
We expect this crossover to be even a smooth process.
Therefore we demand that the stability curve $R_1(R_2)$ is at least
continuously differentiable.
This will fix the parameters $a$ and $R_2^\ast$.
Coles uses the same ansatz $\ell=\alpha a d_n$
but for the whole range $R_2\le 0$~\cite{ColD:67},
which makes the stability boundary discontinuous at $R_2=0$.

Now we address the remaining question at which position $r_p$ the
virtual motion of the rings has to be considered.
At first sight, one might think to choose it as to maximize the virtual 
energy gain.
This leads to $r_p=r_1$.
But more reasonable seems a position where the fluid is able to move in
radial direction if instability indeed sets in.
For Taylor vortices radial fluid motion is strongest near the gap center.
This suggests choosing 
\begin{equation}\label{eRDef}
	r_p = r_1 + \frac{d}{2} \, \Delta(a \frac{d_n}{d}) .
\end{equation}
Experimental data~\cite{SnyHA:70} indicate that this
might be reasonable also for flow patterns other than Taylor vortices.

Let us first evaluate
the critical Reynolds number for resting outer cylinder
defined as $R_{1,c}=R_1(R_2=0)$.
For $R_2=0$ we have $r_n=r_2$, $\ell=\alpha d$ and
$r_p=(r_1+r_2)/2$,
yielding from (\ref{eMain1})
\begin{equation}\label{eRk}
	R_{1,c}(\eta) = \frac{1}{\alpha^2} \, 
		\frac{(1+\eta)^2}{2\eta\sqrt{(1-\eta)(3+\eta)}} ,
\end{equation}
where all constant factors meant by `$\sim$' 
have been absorbed in $\alpha$.
We fix $\alpha$ by matching $R_{1,c}(\eta)$ to
linear stability theory in the narrow gap limit, because in this limit
$R_{1,c}$ is known very accurately, e.g. 
\cite{TayGI:23,CHA:61,LL-6-E}.
The combination
	$4(1-\eta)R_{1,c}^2/(1+\eta)=T_c$
is the critical Taylor number for $R_2=0$ whose value for $\eta\to 1$ is
	$2/\alpha^4 = T_c \approx 3416$
\cite{CHA:61}.
Thus we have $\alpha=0.1556$ corresponding to $\ell=d/6.427$, a reasonable 
value,
which was also found by the same reasoning in~\cite{ColD:67}.

Since $R_{1,c}$ diverges for $\eta\to 1$ and $\eta\to 0$,
one better rescales $R_1$ and $R_2$ to discuss the stability curve
$R_1(R_2)$ for general radius ratio $\eta$, cf also
\cite{WalJ_TsaS_DPrRC:64}.
We employ ${\cal R}_1 = R_1/R_{1,c}$ and ${\cal R}_2 = R_2/(\eta R_{1,c})$,
which will happen to be useful also in the limit when the Reynolds numbers go 
to infinity.

After eliminating $\alpha=\ell/d$ from (\ref{eMain1}) with (\ref{eRk})
we arrive at the final formula for determining the stability boundary:
\begin{equation}\label{eMain}
	\Bigl({\cal R}_1 - {\cal R}_2 \Bigr)^2
	\;\frac{r_n^2 - r_p^2}{r_p^2} \,
	\frac{(1+\eta)^2}{(1-\eta)(3+\eta)}
	=
	\Delta(a\frac{d_n}{d})^{-4}
	.
\end{equation}

We now calculate $a$ as explained above and get
\begin{equation}\label{eA}
	a(\eta) = (1-\eta)\left[\sqrt{\frac{(1+\eta)^3}{2(1+3\eta)}} - 
\eta\right]^{-1} .
\end{equation}
$a$ increases weakly with $\eta$ from $\approx 1.4$ to $1.6$.
We always took this $a(\eta)$, 
but differences to taking
$a=const$ within this range could hardly be noticed in our following figures.
Since $a d_n$ can also be understood as the 
radial size of a Taylor vortex for counterrotating cylinders,
this range of $a$ can be confirmed by inspecting several figures
in Refs.~\cite{TayGI:23,TagR:94,CHA:61,DRAR:81,GebTh_Gro:93}.

\begin{equation}\label{eRast}
	{\cal R}_2^\ast (\eta)=
		-\frac{\eta^2+4\eta+1}{2(1+\eta)^2}
		\sqrt{\frac{3+\eta}{1+3\eta}}
\end{equation}
is also weakly dependent on $\eta$, ranging
between $-0.75$ and $\approx -0.90$.
The continuous differentiability
of the stability curve ${\cal R}_1({\cal R}_2)$ at $a d_n/d=1$
can not trivially be satisfied.
But it is possible because $\Delta$ enters twice in (\ref{eMain}), 
namely on the rhs. via $\ell$, cf~(\ref{eMain1}) and (\ref{eEllDef}),
and on the lhs. via $r_p$, cf~(\ref{eRDef}). 
The argument of $\Delta$ is the same in both instances.
The curvature of the boundary line has opposite sign to the left and
to the right of ${\cal R}_2^\ast$ for all $\eta$.

Eq. (\ref{eMain}) together with (\ref{eNodalDef}),
(\ref{eDeltaDef})--(\ref{eRk}),
and (\ref{eA}) constitute the main result of this work.
It is an implicit equation determining the full
stability curve ${\cal R}_1({\cal R}_2)$ for all radius ratios $\eta$.
It is easily solvable; analytically in special cases, and numerically in
general.

Although we introduced several parameters, only $\alpha$
has been adjusted to linear stability theory.
The others have been fixed by simple physical arguments
and are not arbitrary fit parameters.

\bigskip\noindent\\{\bf 3. Comparison with data and stability 
theory}\\\smallskip

We now discuss the contents of (\ref{eMain}),
i.e., how the stability boundary globally looks like.

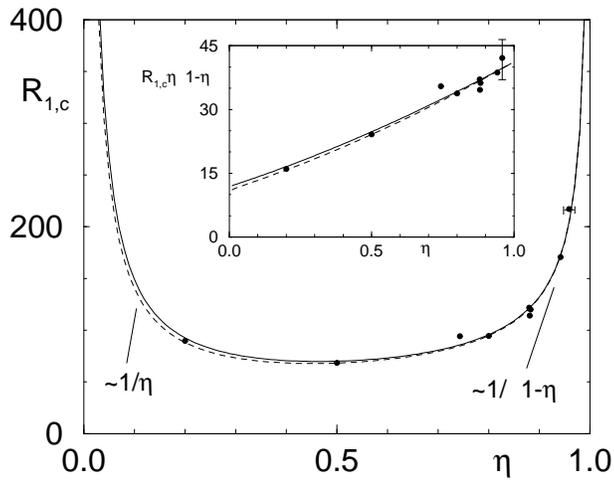
\begin{figure}
\caption[FIG.~1]{
Critical Reynolds number $R_{1,c}$ as a function of $\eta$ from 
Eq.~(\ref{eRk})
(solid line), from linear stability theory~\cite{GebTh_Gro:93} (dashed
line),
and from various experiments
\cite{TayGI:23,DonRJ_FulD:60,ColD:65,SnyHA:68b,AndCD_LiuSS_SwiHL:86}
(symbols).
For the error bar see Fig.~\ref{fEta96}.
Inset: The same for $R_{1,c}$ rescaled to $R_{1,c} \, \eta\sqrt{1-\eta}$.
} \label{fRk}
\end{figure}

Fig.~\ref{fRk} shows $R_{1,c} (\eta)$, Eq.~(\ref{eRk}),
compared with $R_{1,c} (\eta)$ from the Navier--Stokes linear stability 
theory and
some
selected experimental data.
The good overall agreement for $\eta$ approaching $1$ and even for small
$\eta$ comes because Eq.~(\ref{eRk}) has the correct asymptotic behavior,
cf~\cite{TayGI:23,CHA:61} for $\eta\to 1$,
\cite{WalJ_TsaS_DPrRC:64} for $\eta\to 0$, and \cite{GebTh_Gro:93}.
The $1/\eta$ law may be considered as a consequence of our choice
for $r_p$,  since with $r_p=r_1$~\cite{EckB_YaoD:95},
for example, it is $R_{1,c}=const$ for $\eta\to 0$.
Also the prefactor comes out remarkably well.
We find $\eta R_{1,c}=11.9$ in the limit $\eta\to 0$,
whereas $\eta R_{1,c}\approx 11$ from linear stability
theory~\cite{WalJ_TsaS_DPrRC:64}, cf also the inset of Fig.~\ref{fRk}.
Some previous authors gave phenomenological scaling formulae 
which, however, do not show the correct exponents of the asymptotic laws
for $\eta\to 0$ \cite{PRA:31,ColD:67,SnyHA:68b}
or $\eta\to 1$ \cite{SnyHA:68b}.
In particular, Coles finds
	$R_{1,c}\propto 1/(\eta\sqrt{\ln 1/\eta})$
for $\eta\to 0$~\cite{ColD:67}
which does not agree with linear stability theory 
(see the inset of Fig.~1 which clearly does not show a logarithmic 
singularity).
The origin of this logarithmic singularity is that
he took an integral mean of
    $-L^2(r)^\prime/r^3$
over the range $r_1 \le r \le r_n$ for $R_2 < 0$,
and $r_1 \le r \le r_2$ for $R_2 \ge 0$, respectively,
instead of choosing a well-localized position $r_p$ as we do.

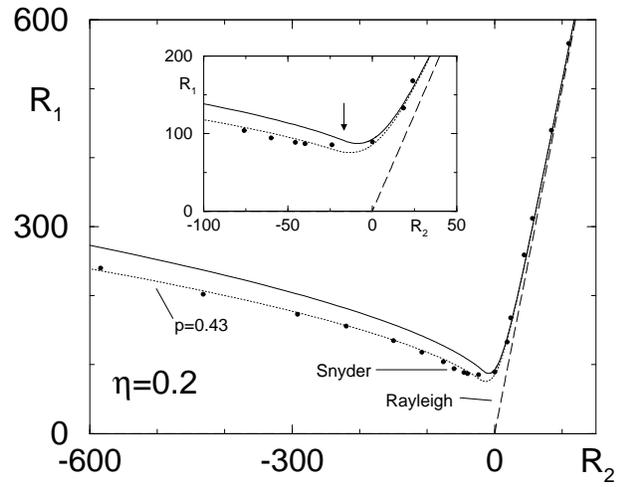
\begin{figure}
\caption[FIG.~2]{
The stability line $R_1(R_2)$ for $\eta=0.2$ from Eq.~(\ref{eMain})
(solid line), and from experiment \cite{SnyHA:68b} (symbols).
The dotted line is our result with the perturbation's position parameter
set to $p=0.43$, see Eq.~(\ref{eRAltDef}).
The arrow marks $R_2^\ast$.
Inset: The same on a magnified scale.
} \label{fEta20}
\end{figure}

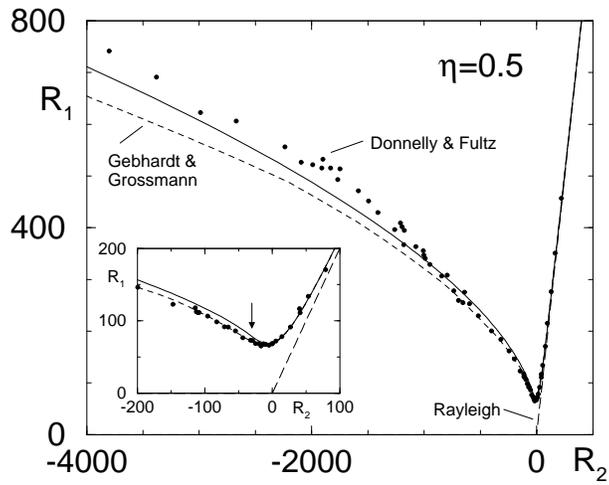
\begin{figure}
\caption[FIG.~3]{
The stability line $R_1(R_2)$ for $\eta=0.5$ from Eq.~(\ref{eMain})
(solid line),
from linear stability theory~\cite{GebTh_Gro:93,GebTh_Gro:pc} (dashed line),
and from experiment \cite{DonRJ_FulD:60} (symbols).
The dashed line may be reproduced with $p=0.46$.
The arrow marks $R_2^\ast$.
Inset: The same on a magnified scale.
} \label{fEta50} 
\end{figure}

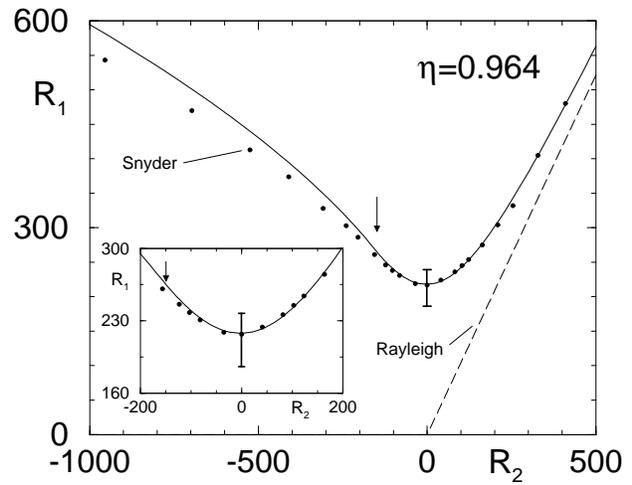
\begin{figure}
\caption[FIG.~4]{
The stability line $R_1(R_2)$ for $\eta=0.964$ from Eq.~(\ref{eMain})
(solid line),
and from experiment \cite{SnyHA:68b} (symbols).
The vertical bar denotes the error in $R_{1,c}$ due to $\eta=0.959\pm 1\%$ 
\cite{SnyHA:68b}.
$p=0.47$ fits the experimental data.
The arrow marks $R_2^\ast$.
Inset: The same on a magnified scale.
} \label{fEta96}
\end{figure}

In Figs.~\ref{fEta20}--\ref{fEta96} the corresponding comparison is offered
for the stability boundary (\ref{eMain}), in the original variables
$R_1(R_2)$,
for $\eta=0.20$, $0.50$, and $0.964$,
covering the whole range for which experimental data seem to be available.
Intermediate values of $\eta$ as in
Refs.~\cite{TayGI:23,ColD:65,SnyHA:68b,AndCD_LiuSS_SwiHL:86}
give essentially the same satisfying picture.

For $R_2 \ge 0$ we find excellent quantitative agreement, 
and for $R_2 < 0$ there still is satisfactory agreement.
The remaining discrepancies may be attributed to several reasons.
First, the experimental data are affected by
the finite length of the cylinders which becomes important for
sufficiently negative $R_2$~\cite{SnyHA:68a,SnyHA:68b}.
This can be clearly seen in Fig.~\ref{fEta50} as a deviation of the
Donnelly--Fultz data from linear stability theory,
the latter being calculated with aspect ratio infinity.
Snyder's data in Figs.~\ref{fEta20} and~\ref{fEta96} are supposed to be free
of finite length effects~\cite{SnyHA:68b}.
Still, the difference of his data to our curve in Fig.~\ref{fEta96} near 
$R_2=0$ is
significantly larger than in other cases
if we take the $\eta$ value reported in \cite{SnyHA:68b},
$\eta=0.959$.
But the deviation is much below the claimed $1\%$ error in the experimental
value for $\eta$~\cite{SnyHA:68b}; we indicated the corresponding error in
$R_{1,c}$ by a bar.
Since the deviation could be caused by the strong $\eta$ dependence
$\sim 1/\sqrt{1-\eta}$, cf Fig.~\ref{fRk},
we match the $R_{1,c}$ values of the experimental data
and of the linear stability theory (and thus our Eq.~(\ref{eRk}))
by choosing $\eta=0.964$.
Now we find the same satisfactory overall picture as for the other $\eta$ 
values.

Our curves correctly show a minimum at some negative $R_2$. This can be
attributed to the crossover in the boundary condition from solid to free. 
$a=1$
leads to a  non-differentiable corner-type minimum at $R_2=0$ and to
considerably higher curves for $R_1(R_2)$ at counterrotation, as was also 
found
in \cite{EckB_YaoD:95}.
Further on, taking $a>1$ and $R_2^\ast=0$ as in \cite{ColD:67}
gives a minimum at $R_2=0$ which is not even continuous.

It is not surprising that our minimum's position
\begin{displaymath}
	{\cal R}_{2,\text{min}}(\eta) =
		-\frac{1-\eta}{2(1+\eta)}
		\sqrt{\frac{3+\eta}{1+3\eta}} ,
\end{displaymath}
its depth and width do not match the measurements too well
since the minimum is located in the crossover region
$R_2^\ast \le R_2 \le 0$. 
We have marked $R_2^\ast$ with an arrow in the figures.
Choosing some other, smoother crossover function $\Delta$ in
(\ref{eDeltaDef})
allows to modify the location and the width of the minimum.

Interestingly,
for $\eta\to 1$ the minimum is exactly at $R_2=0$, cf Fig.~\ref{fEta96},
independent of the value
of $a$ (as long as $a>1$) and of the precise form of $\Delta$,
and nicely compatible with experiment \cite{TayGI:23,SnyHA:68b}
and theory \cite{TayGI:23,CHA:61,WalJ_TsaS_DPrRC:64,DRAR:81,GebTh_Gro:93}.

For $R_2$ far left of
the minimum our curves calculated from (\ref{eMain}) are clearly 
systematically too high.
Before discussing this,
we complete our analysis by looking at interesting special cases,
in which explicit formulae follow.

\bigskip\noindent\\{\bf 4. Asymptotic formulae}\\\smallskip

As long as the vortex flow fills the whole gap, i.e.,
for ${\cal R}_2 \ge {\cal R}_2^\ast$, we have $\Delta(a d_n/d)=1$;
then Eq.~(\ref{eMain}) can be solved for ${\cal R}_1$:
\begin{eqnarray}
	{\cal R}_1 &=& \frac{1-\eta}{3+\eta} \, {\cal R}_2 +
		\sqrt{1+\left( 2\,\frac{1+\eta}{3+\eta}\,{\cal 
R}_2\right)^2},\\
	&& \text{for} \;
	{\cal R}_2 \ge {\cal R}_2^\ast .\nonumber
\end{eqnarray}
This implies a leading correction $\propto{\cal R}_2^{-2}$  relative to the
Rayleigh line ${\cal R}_1 = {\cal R}_2$ for large ${\cal R}_2$.

For ${\cal R}_2 < {\cal R}_2^\ast$, the quantity
\begin{equation}\label{eEps}
	\varepsilon 
	= \bigl( \frac{r_n}{r_1} \bigr)^2 - 1
	=  \frac{1-\eta^2}{\eta^2} \, \frac{{\cal R}_1}{{\cal R}_1 - {\cal 
R}_2}
\end{equation}
is a suitable parameter for discussing the strongly counterrotating case as
well as the narrow and wide gap limits.
For $\varepsilon\to 0$,
corresponding to $r_n \to r_1$,
we find from (\ref{eMain})
\begin{eqnarray}\label{eMinusInfty}
	-\eta^2{\cal R}_2 & = & 
		\Bigl( \frac{{\cal R}_1}{C} \Bigr)^{5/3} \!\! - \eta^2{\cal 
R}_1 
		, \;\, {\cal R}_1 \gg (1\!-\!\eta^2)^{3/2} C^{5/2}, \\
	C (\eta) & = & \left( \frac{2^5 (3+\eta)}
			{a^4(2-a)(1+\eta)^7} \right)^{1/5}.
	\nonumber
\end{eqnarray}
The leading term for large $|{\cal R}_2|$ reads 
\begin{equation}\label{eLeadMinusInfty}
	{\cal R}_1 = C(\eta) \, |\eta^2 {\cal R}_2|^{3/5},
	\quad {\cal R}_2 \to -\infty.
\end{equation}
The exponent $3/5$
was already found by Donnelly and Fultz~\cite{DonRJ_FulD:60}
by a dimensional argument.
That, of course, did not yet give the complete $\eta$
dependent prefactor $\eta^{6/5} C(\eta)$ and also not the non-trivial 
corrections
contained in (\ref{eMinusInfty}).
These turn out to be non-neglegible in the range of ${\cal R}_2$ accessible
to experiment.
Since $\varepsilon \propto |\eta^2{\cal R}_2|^{-2/5}$ for sufficiently 
large $|\eta^2 {\cal R}_2|$,
the convergence 
to the leading behavior (\ref{eMinusInfty}) and (\ref{eLeadMinusInfty})
is rather bad in terms of ${\cal R}_2$
for fixed $\eta$ and becomes worse with decreasing $\eta$.

Coles~\cite{ColD:67} has chosen the value of $a$ by matching 
his analytic formula to linear stability theory for $\eta\to 1$ in the limit
$R_2\to -\infty$.
This choice leads to an $\eta$ independent value of $a\approx 1.3$.
This means he adjusted his formula to two values from linear stability theory
whereas we only use one.

On the other hand, for $\eta\to 1$ Eq.~(\ref{eMinusInfty}) becomes exact for
all ${\cal R}_2 \le {\cal R}_2^\ast$.
In this limit we have ${\cal R}_2^\ast = -3/4$ and $a=8/5$
from (\ref{eRast}) and (\ref{eA}).
Furthermore, the position of the minimum approaches ${\cal R}_{2,\text{min}} 
= 0$,
and the stability boundary becomes symmetric for
$|{\cal R}_2| \le |{\cal R}_2^\ast|$, cf Fig.~\ref{fEta96}.

Therefore,
in Fig.~\ref{fEta96} the asymptotic formula (\ref{eMinusInfty})
would be indistinguishable from the full curve,
whereas in Figs.~\ref{fEta20} and~\ref{fEta50}
it would be a rather poor approximation.

Of particular interest is the wide gap limit $\eta\to 0$,
since it has not yet been examined very much.
It is accompanied by large $\varepsilon$ for any finite ${\cal R}_2$ 
interval,
cf~(\ref{eEps}).
For $\varepsilon\to\infty$ the relevant expansion parameter in (\ref{eMain}) 
is
$\varepsilon^{-1/2}$.
We keep the two leading terms and consider $\eta$ small in them:
\begin{eqnarray}\label{eEtaZero}
	{\cal R}_1 &=& \frac{1}{2}\sqrt{3} +
		\frac{3+2\sqrt{2}}{1+\sqrt{2}} \, 
			\eta\,\sqrt{1 - \frac{2}{\sqrt{3}} \, {\cal R}_2} ,\\
	&&\text{for}\;
		-{\cal R}_1/\eta^2 \ll {\cal R}_2 \le {\cal 
R}_2^\ast.\nonumber
\end{eqnarray}
Here ${\cal R}_2^\ast = -\sqrt{3}/2$ and $a=\sqrt{2}$ for $\eta\ll 1$,
cf~(\ref{eRast}) and (\ref{eA}).
Because of the bad convergence in terms of $\varepsilon^{-1/2}$ to this 
limiting behavior
the line ${\cal R}_1({\cal R}_2)$ according to (\ref{eEtaZero}) is clearly 
visible
only for $\eta\lesssim 0.01$.
Of course, for fixed $\eta >0$ and sufficiently largely negative ${\cal R}_2$
equation (\ref{eMinusInfty}) is valid again because then $\varepsilon \to 0$ 
instead
of $\to\infty$.
As can be seen from (\ref{eLeadMinusInfty}) and (\ref{eEtaZero}) the most 
reasonable
variables for strongly counterrotating cylinders, i.e.,
${\cal R}_2\to -\infty$,
are ${\cal R}_1$ and $\eta^2 {\cal R}_2$.
Note that the meaning of their ratio is
$\eta^2{\cal R}_2/{\cal R}_1 = \omega_2/\omega_1$.
This frequency ratio as a parameter has already been used in
\cite{TayGI:23,CHA:61}.

\bigskip\noindent\\{\bf 5. Improvements}\\\smallskip

Now we discuss possible corrections to our main result.
It turns out that the precise value of $r_p$ has a surprisingly well 
observable
influence on the stability curve.

To see this, we generalize (\ref{eRDef}) to
\begin{equation}\label{eRAltDef}
	r_p = r_1 + p \, d \, \Delta(a \frac{d_n}{d})
\end{equation}
with the new parameter $p$ describing the position of the disturbance;
$p=1/2$ gives the previous case, cf (\ref{eRDef}),
to locate it in the center of the relevant gap.

Treating $p$ as a free fit parameter
we find the remaining discrepancies between our stability boundary
and that of experiment or linear stability calculations nearly
removed for $p=0.47, 0.46, 0.43$, corresponding to $\eta=0.964, 0.5, 0.2$,
respectively.
This is exemplified in Fig.~\ref{fEta20} by the dotted line;
for the other $\eta$ values the agreement is similarly good.
Position shifts of this magnitude are confirmed by several figures in
Refs.~\cite{TayGI:23,CHA:61,DRAR:81,GebTh_Gro:93}
which show that the center of each Taylor vortex is situated nearer towards 
the inner cylinder.
Analytically we find roughly
$a\sim 1/p$ and $C$ approximately $\sim p$ if $p$ is near $1/2$ and for 
$\eta$ not to
small.
We could not figure out a simple physical condition for fixing $p(\eta)$.
For example, second order continuous differentiability of the stability
boundary at ${\cal R}_2^\ast$ turns out not to be realizable.

Our argument does not take into consideration non-axisymmetric perturbations
of the laminar flow.
As we know from experiment~\cite{SnyHA:70}
and linear stability theory~\cite{DRAR:81},
the relevant perturbations are indeed non-axisymmetric beyond a finite 
negative
$R_2$.
They lead to a somewhat lower stability boundary than that determined with
axisymmetric disturbances.
The difference for the values of $R_2$ considered here is pretty small.
It turns out that our deviations from the correct stability boundary are of
the same order of magnitude.
Furthermore these deviations start already near $R_{2,\text{min}}$.
We therefore think it to be acceptable that the possibility of 
non-axisymmetric
disturbances is neglected.

Of course, also other corrections have to be considered.
For example, energy dissipation is curvature dependent.
Therefore it seems natural to extend the rhs.\ of (\ref{eDissipation}) to
	$(\nu^2/\ell^4)(\delta r)^2 (1+b \ell/r_p) $
with $b$ a constant.
$\ell/r_p$ vanishes for $\eta\to 1$ or for ${\cal R}_2\to -\infty$
and is rather small even for $\eta\to 0$;
for example in the case ${\cal R}_2 = 0$
we have $\ell/r_p=\alpha\approx 0.16$.
Thus (\ref{eMinusInfty}) remains valid, but with a different $a$ in $C$.
Note, without this curvature correction the value for $a(\eta)$, 
Eq.~(\ref{eA}), is close to
$a=8/5$ which minimizes $C$ for fixed $\eta$ but variable $a$.
In general, this means the curvature correction shifts the stability boundary 
{\em up\/}wards
if ${\cal R}_2$ is well below ${\cal R}_2^\ast$
counteracting the influence of the $r_p$ shift.

\bigskip\noindent\\{\bf 6. Summary}\\\smallskip

Similar arguments as are presented here for Taylor--Couette flow have been
invoked to elucidate the physical mechanism of the
Ray\-leigh--B{\'e}nard instability~\cite{NorC_PomY_VelMG:77}.
But in the Taylor--Couette system it carries much further since 
this has two additional parameters,
$\eta$ and $R_2$, whose influence on the shape of the stability boundary we 
could successfully
explain.
At Rayleigh--B{\'e}nard instability
$a$ corresponds to the ratio of the $\alpha$'s for 
the two distinct cases of free and solid boundary conditions.
From~\cite{CHA:61} we find for this ratio $\approx 1.13$.
We understand this value closer to $1$ than we found here for the 
counterrotating
Taylor--Couette system
from the stronger influence of the surface tension in an open
Rayleigh--B{\'e}nard system in contrast to the completely missing surface
tension at the nodal surface $r_n$ in Taylor--Couette flow.
Therefore this nodal surface is much softer, allowing for a larger extension 
of
the Taylor vortices beyond $r_n$.
One could check this by putting another fluid layer on top of the original
Rayleigh--B{\'e}nard fluid;
if this diminishes the surface tension, a larger $a$ should result.

To conclude:
While accurate results for the onset of the Taylor--Couette instability
derived by linear stability theory are available, 
in this paper we have drawn a picture that
gives comprehensive physical insight into the instability mechanism
and yields simple analytical expressions for the stability boundary in the
full parameter space.
These agree astonishingly well with experimental data and with the
rigorous theory,
especially in regard of how straightforward the argument is.
The position where the initial perturbation occurs
and the change of the type of boundary when the nodal surface moves inwards
turned out to be important for the wide gap behavior and for the existence of 
a minimum of the stability boundary for negative $R_2$.
The values found for the parameters $\alpha$ and $a$ are reasonable and
agree with experiment as well.
Assuming a radius ratio dependent small shift of the perturbation's initial 
position
towards the inner cylinder we were able to understand the remaining 
quantitative discrepancies.

We believe that these simple ideas
should be applicable to instabilities in more complicated
rotating fluid flows, where the standard linear stability
analysis is too difficult.

{\bf Acknowledgments}:
We thank Thomas Gebhardt and Martin Holthaus for their interest and help. 
S. G. acknowledges stimulating discussions about similar analysis of thermal
convection with Stefan Thomae, quite some time ago, during a sabbatical stay 
at the
Max-Planck-Institut f{\"u}r Mathematik in Bonn.


\end{document}